\documentclass[a4paper,aps,preprintnumbers,showpacs,superscriptaddress,nofootinbib,amsmath,amssymb]{revtex4}
\usepackage{graphicx}
\usepackage{hyperref}
\usepackage{subfigure}
\usepackage{multirow}
\usepackage[toc,page]{appendix}

\def\imo{i}

\begin{document}
\title{Quasinormal modes of the four-dimensional black hole in Einstein-Weyl gravity}
\author{A. F. Zinhailo}\email{F170631@fpf.slu.cz}
\affiliation{Institute of Physics and Research Centre of Theoretical Physics and Astrophysics, Faculty of Philosophy and Science, Silesian University in Opava, CZ-746 01 Opava, Czech Republic}

\begin{abstract}
In this paper we consider quasi-normal modes of test scalar and electromagnetic fields in the background of a non-Schwarzschild black hole in the Einstein-Weyl gravity. We find the dependence of the modes on the new parameter $p$, which is related to the coupling constant  $\alpha$, for different values of the angular parameter $l$ and show that when $l$ is increasing, the frequencies tend to linear dependence on $p$. We have also examined the case of a massive scalar field for which we have shown that the arbitrarily long lived modes (called quasi-resonances) exist in the spectrum.

\end{abstract}
\maketitle

\section{Introduction}
As is known, the general theory of relativity leaves open some fundamental questions, such as: the singularity problem, construction of non-contradictory quantum gravity, the problems of dark matter and dark energy. The search for answers to these questions led to the emergence of new alternative theories. Some of these alternative theories are inspired by string theory, which suggests its ways of resolution of some of the above problems. One of the most popular approaches is related to adding higher than the first order in curvature terms to the Einstein action. If no other matter fields are included for four-dimensional space-time $D=4$, the most general theory up the second order in curvature is described by the following Lagrangian:
%
\begin{eqnarray} \label{lagranzian}
L &=& \sqrt{-g} (\gamma R - \alpha C_{\mu\nu \rho\sigma} C^{\mu\nu \rho\sigma} + \beta R^2),
\end{eqnarray}
here $g$ is determinant of the matrix of a metric tensor; $\alpha$, $\beta$ and $\gamma$ are constants; $C_{\mu\nu \rho\sigma}$ is the Weyl tensor. This theory has been recently considered in a number of papers \cite{Einstein-Weyl:2018pfe}, \cite{QuadraticGravity:2016kip}.

On the other hand, in the last time interest in studying perturbations of black holes is growing \cite{Kokkotas:1999bd}, \cite{Berti:2009kk}, \cite{Konoplya:2011qq}. This was greatly contributed by observations of LIGO and VIRGO collaborations of the gravitational-wave signal at the  merger of two black holes  \cite{Abbott:2016blz}. The results of their observations are in concordance with General Relativity. The value ​​for modes are given within a three percent error. However, the angular momentum and mass of a newly formed black hole are determined from the conservation laws and have significant uncertainty in tens of percent. This makes it possible to obtain close values for dominant quasinormal frequencies while considering two black holes with different masses and momenta in two different theories of gravity. It leaves the window for alternative theories wide open \cite{Konoplya:2016pmh}, \cite{Yunes:2016jcc}, \cite{Berti:2018vdi}. This enhanced consideration of various alternative theories of gravity, including theories with higher curvature corrections.

Perturbations of black holes in theories with higher curvature corrections have been recently studied in a number of papers when considering higher dimensional black holes in the context of gauge/gravity duality and higher dimensional gravity \cite{dualityAndHigherDimensionalGravity:2004xx}, in the four-dimensional Einstein-dilaton-Gauss-Bonnet theory \cite{Einstein-dilaton-Gauss-Bonnet:2016enn}, in the Chern-Simons gravity \cite{Gonzalez:2010vv} and in the quadratic gravity \cite{QuadraticGravity:2016kip}, \cite{Cai:2015fia}, \cite{Lu:2015cqa}, \cite{Lu:2015psa}. None of the these alternative theories of gravity is discarded by the LIGO and VIRGO observations because of the above mentioned uncertainty in measured mass and angular momentum. Here we will consider the Einstein-Weyl theory of gravity. Apart from its generality, the latter quadratic theory is interesting also because, unlike many other theories of gravity, it allows for two asymptotically flat solutions representing a black hole: one is the usual Schwarzschild solution and the other is essentially non-Schwazrschild solution. The non-Schwarzschild solution has a remarkable property: when the coupling constant approaches its extremal value, the black hole approaches the massless state, having a non-zero radius at the same time. Thus,  near extremal alternatives to the Schwarzschild solution are a kind of ``light black holes'' which are qualitatively different from its Einstein analogues. Once quasinormal modes in various alternative theories of gravity are found, the comparison of theoretical predictions with the experimental data by current and future LIGO and VIRGO experiments would help to constrain the parameters in those theories. Thus, it is important to accumulate data on quasinormal modes of black holes in various theories of gravity.

Although there is some literature on quasinormal modes on scalar field in the background of Einstein-Weyl black holes, the accurate calculations of quasinormal modes have not yet been done. The eikonal formula for quasinormal modes of test fields was obtained for the black hole in this theory in \cite{Kokkotas:2017zwt}. However, no analysis of the fundamental quasinormal modes was performed there. At the same time, the analysis of the fundamental mode suggested in \cite{Cai:2015fia} only for a massless test field suffers, in our opinion, from numerical inaccuracy. In this paper we investigate the quasinormal spectrum of a scalar and a electromagnetic test fields and find the analytic expressions for the quas-normal modes for different values of the multipole number $l$. We also consider the case of a massive scalar field. The interest of this case is connected with the possibility of a existence of undamped oscillations, i. e. quasiresonance, for sufficiently large masses of the field. First this phenomenon was observed for the charged field of the Reissner-Nordstr\"{o}m metric in \cite{Ohashi:2004wr} numerically and in \cite{Hod:2016jqt}, \cite{Hod:2017gvn} analytically. This phenomenon was further studied for Schwarzschild and Kerr cases in \cite{Konoplya:2006br}, \cite{Konoplya:2004wg}. Moreover, recently in \cite{Konoplya:2018qov} it was shown that the quasi-resonances exist even at non-minimal coupling of the field, that is, they are not artifacts of the minimal coupling. Here we have shown the existence of a quasi-resonance for the massive scalar field in the background of the non-Schwarzschild solution in the quadratic gravity.

The structure of this work is as follows. In Sec.~\ref{sec:metricsection} we consider the basic concepts, introduce a metric and a wave equation for a spherically symmetric black hole. In Sec.~\ref{sec:wkbmethod}, the basic principles of the WKB method are briefly considered, quasi-normal modes for test scalar and electromagnetic fields of massless (Sec.~\ref{sec:massless}) and massive (Sec.~\ref{sec:mass}) are found, a comparative analysis is made for the result obtained for the scalar field with the results in other studies. The possibility of the existence of quasiresonance is also shown.

\section{Black hole metric and Analytics for the wave equation}\label{sec:metricsection}
The metric for a spherically symmetric black hole was found in \cite{Kokkotas:2017zwt} and in general case is:
%
\begin{eqnarray}\label{metric}
dS^2 &=&
-A(r)dt^2+\dfrac{dr^2} {B(r)}+r^2 (\sin^2 \theta d\phi^2+d\theta^2).
\end{eqnarray}

The metric functions $A(r)$ and $B(r)$ were obtained numerically in \cite{Lu:2015cqa}, \cite{Lu:2015psa} and an analytical approximation was found for this numerical solution in \cite{Kokkotas:2017zwt} and have the forms:

\begin{subequations}
\begin{eqnarray}
A(r)&=&
[(r-r_0) (152124199161 (873828 p^4-199143783 p^3+806771764 p^2-1202612078 p+604749333) r^4
\nonumber \\&&
+78279 (1336094371764 p^6-300842119184823 p^5+393815823540843 p^4+2680050514097926 p^3
\nonumber \\&&
-9501392159249689 p^2+10978748485369369 p-4249747766121792) r_0 r^3-70372821 (1486200636 p^6
\nonumber \\&&
+180905642811 p^5+417682197141 p^4-1208134566031 p^3-324990706209 p^2+3382539200269 p
\nonumber \\&&
-2557857695019) r_0^2 r^2+(-104588131327314156 p^6+23549620247668759617 p^5+435688050031083222417 p^4
\nonumber \\&&
-2389090517292988952355 p^3+3731827099716921879958 p^2-2186684376605688462974 p
\nonumber \\&&
+389142952738481370396) r_0^3 r+31 (3373810687977876 p^6+410672271594465801 p^5
\nonumber \\&&
-14105000476530678231 p^4+51431640078486304191 p^3-71532183052581307042 p^2+43250367615320791700 p
\nonumber \\&&
-9476049523901501640) r_0^4)] [152124199161 r^3 ((873828 p^4-199143783 p^3+806771764 p^2-1202612078 p
\nonumber \\&&
+604749333) r^2-2 (873828 p^4-47583171 p^3+386036980 p^2-678598463 p+341153481t) r_0 r+899 (972 p^4
\nonumber \\&&
+115659 p^3-38596 p^2-1127284 p+1101579) r_0^2)]^{-1},
\end{eqnarray}
\begin{eqnarray}
B(r)&=&
[215672004025 (r-r_0) ((3251230164 p^3-14548777134 p^2+20865434326 p+23094914865) r^2
\nonumber \\&&
+(-6502460328 p^3+52856543928 p^2-100077612184 p+32132674695) r_0 r+6 (541871694 p^3-6384627799 p^2
\nonumber \\&&
+13202029643 p+2626009760) r_0^2){}^2 (152124199161 (873828 p^4-199143783 p^3+806771764 p^2-1202612078 p
\nonumber \\&&
+604749333) r^4+78279 (1336094371764 p^6-300842119184823 p^5+393815823540843 p^4+2680050514097926 p^3
\nonumber \\&&
-9501392159249689 p^2+10978748485369369 p-4249747766121792) r_0 r^3-70372821 (1486200636 p^6
\nonumber \\&&
+180905642811 p^5+417682197141 p^4-1208134566031 p^3-324990706209 p^2+3382539200269 p
\nonumber \\&&
-2557857695019) r_0^2 r^2+(-104588131327314156 p^6+23549620247668759617 p^5+435688050031083222417 p^4
\nonumber \\&&
-2389090517292988952355 p^3+3731827099716921879958 p^2-2186684376605688462974 p
\nonumber \\&&
+389142952738481370396) r_0^3 r+31 (3373810687977876 p^6+410672271594465801 p^5
\nonumber \\&&
-14105000476530678231 p^4+51431640078486304191 p^3-71532183052581307042 p^2+43250367615320791700 p
\nonumber \\&&
-9476049523901501640) r_0^4)] [(152124199161 r ((873828 p^4-199143783 p^3+806771764 p^2-1202612078 p
\nonumber \\&&
+604749333) r^2-2 (873828 p^4-47583171 p^3+386036980 p^2-678598463 p+341153481) r_0 r+899 (972 p^4
\nonumber \\&&
+115659 p^3-38596 p^2-1127284 p+1101579) r_0^2) (-464405 (3251230164 p^3-14548777134 p^2+20865434326 p
\nonumber \\&&
+23094914865) r^3+464405 (6502460328 p^3-52856543928 p^2+100077612184 p-32132674695) r_0 r^2
\nonumber \\&&
+(1244571650887908 p^3+17950319416564777 p^2-53210739821255918 p+5097428297648940) r_0^2 r
\nonumber \\&&
-635371 (4335198168 p^3-42352710803 p^2+90235778452 p-49464019740) r_0^3){}^2]^{-1}.
\end{eqnarray}
\end{subequations}

We return to equation (\ref{lagranzian}). As mentioned above, in the framework of the quadratic gravity, we should have two analytical solutions: the usual Schwarzschild solution and a solution with the quadratic correction. As shown in \cite{Lu:2015cqa}, \cite{Lu:2015psa}, we can put $\beta=0$. We also set $\gamma=1$. The remaining undefined constant $\alpha$ we rewrite through the dimensionless parameter $p$, which parameterizes the solutions up to the rescaling and it  can be determined through $\alpha$ as follows:
\begin{equation}
p=\dfrac{r_0}{\sqrt{2 \alpha}},
\end{equation}
where $r_0$ is a black hole radius. The Schwarzschild metric is valid for all $p$, however the second solution for an asymptotically flat black hole appears at some intervals of the values of this parameter, for small values \cite{Lu:2015cqa}, \cite{Lu:2015psa}. The approximate minimum and maximum values of $p$ are:
%
\begin{subequations}
\begin{eqnarray}
p_{min}\approx\dfrac{1054}{1203}\approx0.876,\label{p_min}\\
p_{max}\approx1.14.
\end{eqnarray}
\end{subequations}

We will consider quasi-normal modes of test fields, scalar and electromagnetic, in the background of the black hole. Their perturbations can be represented in terms of the general relativistic Klein-Gordon equation:
\begin{equation}\label{Klein-Gordon}
(\Box-\mu^2) \Phi= 0,
\end{equation}
for a massive scalar field equation (\ref{Klein-Gordon}) has the form:
\begin{equation}\label{gensc}
\dfrac{1}{\sqrt{-g}} \partial_\mu(\sqrt{-g} g^{\mu\nu} \partial_\nu\Phi)-\mu^2 \Phi = 0,
\end{equation}
for an electromagnetic field the general covariant Maxwell equations read as follows:
\begin{equation}\label{genelm}
\dfrac{1}{\sqrt{-g}} \partial_\mu(F_{\rho\sigma}g^{\rho\nu}g^{\sigma\mu}\sqrt{-g})=0.
\end{equation}

Here $A_{\mu}$ is a vector potential and $F_{\rho\sigma}=\partial_\rho A^\sigma - \partial_\sigma A^\rho$. The function $\Phi$ can be represented in terms of spherical harmonics and radial part as follows:
\begin{equation} \label{psi}
\Phi(t,r,y,\phi)=e^{\pm\imo\omega t}R_{\omega l}(r)Y_l(\theta,\phi).
\end{equation}

Here $R_{\omega l}(r)$ is a radial part, $Y_l(\theta,\phi)$ is a usual spherical harmonics, $l$ is a angular harmonic index. For convenience, we rewrite the radial component $R_{\omega l}(r)$ through the new function:
\begin{equation}
\Psi_{\omega l}(r)=r R_{\omega l}(r).
\end{equation}

Thus, taking into account the replacement (\ref{psi}), equations (\ref{gensc}) and (\ref{genelm}) in terms of the ``tortoise coordinate'' $r_*$ can be represented in the general form:
\begin{equation}  \label{klein-Gordon}
\dfrac{d^2 \Psi}{dr_*^2}+(\omega^2-V(r))\Psi=0,  \qquad dr_*=\dfrac{dr}{\sqrt{A(r) B(r)}}.
\end{equation}

We will consider test scalar and electromagnetic fields. The effective potentials for the general metric (\ref{metric}) can be written in the forms the following forms \cite{Konoplya:2006rv}:
%
\begin{subequations}\label{potentials}
\begin{eqnarray}\label{scalarpotential}
V_{s}(r)=\dfrac{l(l+1)A(r)}{r^2}+\dfrac{B(r)\dfrac{\partial A}{\partial r}+A(r)\dfrac{\partial B}{\partial r}}{2r},\\
V_{e}(r)=\dfrac{l(l+1)A(r)}{r^2}.
\end{eqnarray}
\end{subequations}

\section{Quasi-normal modes for test fields}\label{sec:scalarsection}

\subsection{The WKB method} \label{sec:wkbmethod}
The solution of equation (\ref{klein-Gordon}) can be found with the help of the WKB approximation \cite{Schutz:1985zz}. This approach has application in quantum mechanics for finding the first terms in the expansion of the wave function, provided that the amplitude or frequency of the oscillation changes smoothly. When solving the wave equation, strictly fixed boundary conditions are imposed: for a functions $\Psi$ there are only incoming waves at the horizon ($r_*\rightarrow-\infty$) and only the outgoing waves at the infinity ($r_*\rightarrow+\infty$). Finally, we obtain a complex number for the frequency $\omega$, in which the real part is the real oscillation frequency, and imaginary part is the damping rate of oscillations in terms of the parameters of the black hole. It is known that the WKB series converges only asymptotically. Nevertheless, it is observed in many papers \cite{Iyer:1986np}, \cite{Matyjasek:2017psv}, \cite{Konoplya:2003ii} that higher WKB orders usually produce more accurate results. In our work we will use the sixth-order wkb-corrections obtained in  \cite{Konoplya:2003ii}:
\begin{equation}\label{wkb}
\dfrac{i(\omega^2-V_0)}{\sqrt{-2 V_0''}}-\sum\limits_{i = 2}^{6}\Lambda_i=n+\dfrac{1}{2}.
\end{equation}

Here, the $\Lambda_i$  are the correction term of the i-th order and $\Lambda_i$ depend on the value of the potentials $V$ and its derivative at the maximum and are explicitly represented in \cite{Iyer:1986np}, \cite{Konoplya:2003ii}; $ n=0,1,2,..$ is a overtone number. It is sufficient to use the 6th order WKB formula for small $n$, while $n \leq l$. Here we are limited only by the fundamental mode $n=0$, because the contribution of higher overtones in the resultant signal is negligible \cite{Konoplya:2011qq}. In addition, since gravitational waves are very weak, only the fundamental mode of the spectrum can be extracted from a signal with sufficient accuracy.

\subsection{Quasi-normal modes for massless fields}\label{sec:massless}

We find the quasi-normal modes for given potentials (\ref{potentials}) with the help of the sixth order WKB formula \cite{Konoplya:2006rv} with the values of the multipole number $l=0,1,2$ for the scalar field ($s=0$) and $l=1,2,3$ for the electromagnetic field ($s=1$). The results are presented in table \ref{tab:tableone}.

\begin{figure}[ht]
\vspace{-4ex} \centering \subfigure[]{
\includegraphics[width=0.29\linewidth]{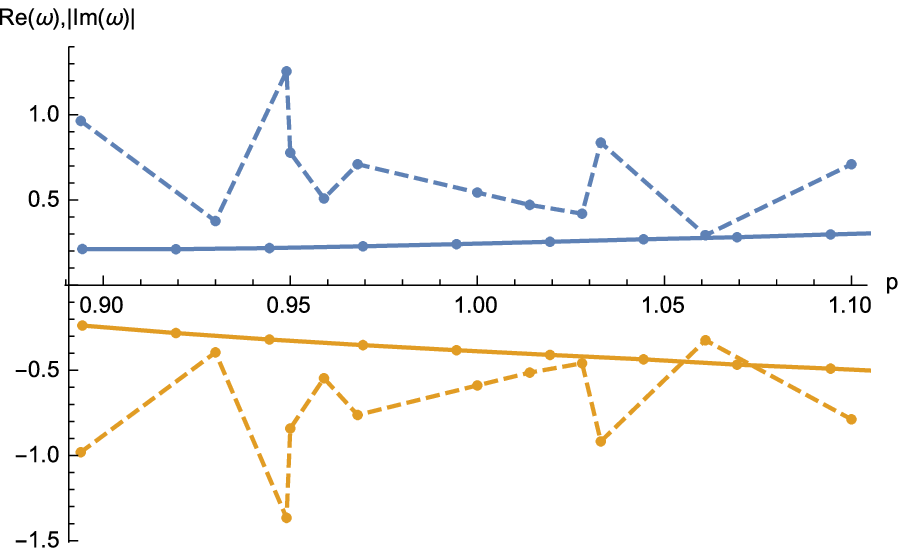} \label{fig:actuatorscouplingSheme_a} }
\hspace{4ex}
\subfigure[]{
\includegraphics[width=0.29\linewidth]{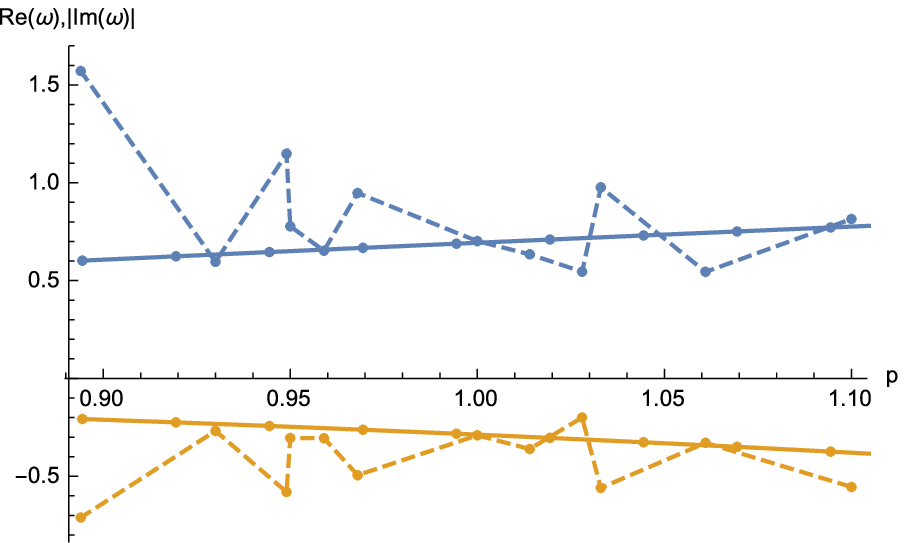} \label{fig:actuatorscouplingSheme_b} }
\hspace{4ex}
\subfigure[]{ \includegraphics[width=0.29\linewidth]{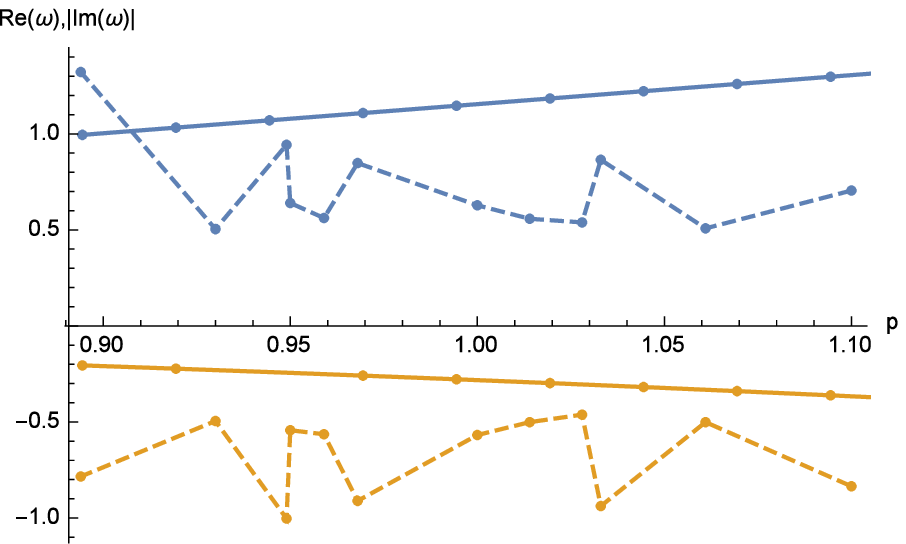} \label{fig:actuatorscouplingSheme_c} }
\caption{The fundamental quasinormal mode ($n=0$) for the scalar field ($s=0$), blue line is real part of frequencies, red line is imaginary part; dashed lines are the data obtained in the work \cite{Cai:2015fia}, positive values for the real part and negative values for the imaginary part:
 \subref{fig:actuatorscouplingSheme_a} , $l = 0$;
 \subref{fig:actuatorscouplingSheme_b} , $l = 1$;
 \subref{fig:actuatorscouplingSheme_c} , $l = 2$.} \label{fig:threeDMcases}
 \label{ris:one}
\end{figure}

\begin{figure}[ht]
\vspace{-4ex} \centering \subfigure[]{
\includegraphics[width=0.29\linewidth]{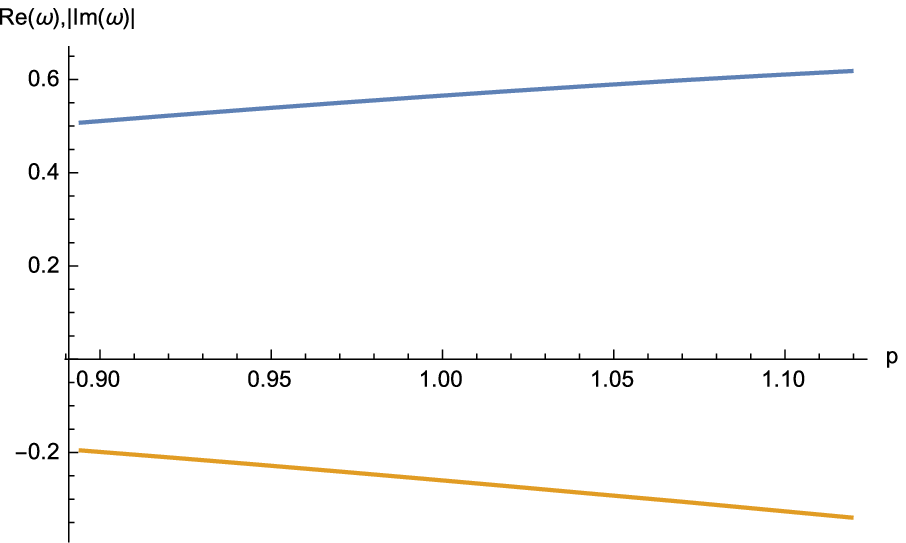} \label{fig:actuatorscouplingSheme_aa} }
\hspace{4ex}
\subfigure[]{
\includegraphics[width=0.29\linewidth]{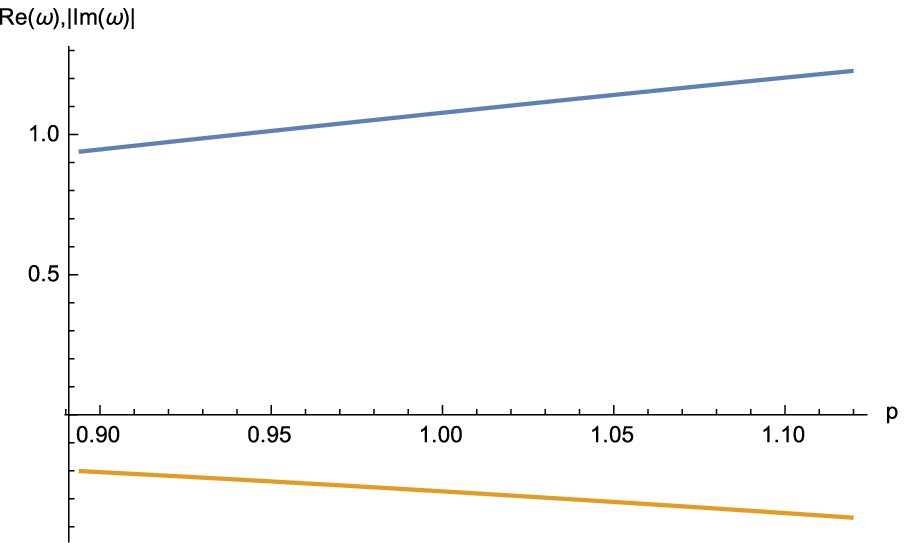} \label{fig:actuatorscouplingSheme_bb} }
\hspace{4ex}
\subfigure[]{ \includegraphics[width=0.29\linewidth]{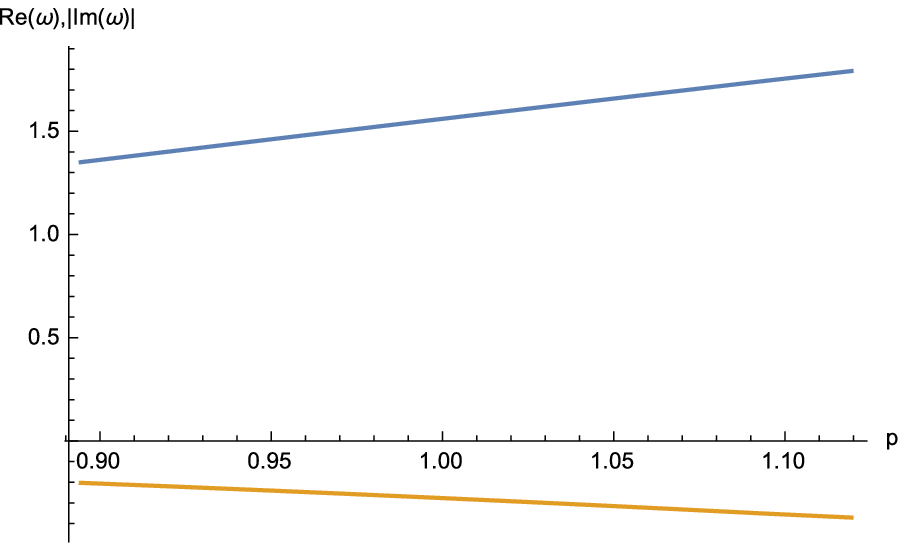} \label{fig:actuatorscouplingSheme_cc} }
\caption{The fundamental quasinormal mode ($n=0$) for the electromagnetic field ($s=1$), blue line is real part of frequencies, orange line is imaginary part:
 \subref{fig:actuatorscouplingSheme_aa} , $l = 1$;
 \subref{fig:actuatorscouplingSheme_bb} , $l = 2$;
 \subref{fig:actuatorscouplingSheme_cc} , $l = 3$.}
 \label{ris:two}
\end{figure}

From the values obtained, we construct the dependencies of the real and, separately, imaginary parts of the frequency $\omega$ on the values of the parameter $p$ for various multipole numbers $l$. As can be seen in figures \ref{ris:one} and \ref{ris:two}, the dependence $\omega(p)$ tends to be linear. Thus, it is possible to obtain approximate laws of variation of the imaginary and real parts as functions of $p$ for different values of the angular number $l = 0$,  $l = 1$,  $l = 2$. From the above data one can see that the non-Schwarzschild branch of the Einstein-Weyl theory is always characterized by higher oscillation frequency and quicker damping of the fundamental mode.
\begin{subequations}\label{scalarline}
\begin{eqnarray}
Re(\omega_{s=0,l=0})\approx-0.228+0.477 p,
\nonumber\\
Im(\omega_{s=0,l=0})\approx0.680-1.070p;\\
Re(\omega_{s=0,l=1})\approx-0.151+0.84314 p,
\nonumber\\
Im(\omega_{s=0,l=1})\approx0.562- 0.854 p;\\
Re(\omega_{s=0,l=2})\approx-0.359+1.514 p,
\nonumber\\
Im(\omega_{s=0,l=2})\approx0.509-0.794 p.
\end{eqnarray}
\end{subequations}

Accordingly, for the electromagnetic field (FIG. 2) $l = 1$,  $l = 2$,  $l = 3$, we have:
\begin{subequations} \label{elmline}
\begin{eqnarray}
Re(\omega_{s=1,l=1})\approx0.070 + 0.493 p,
\nonumber\\
Im(\omega_{s=1,l=1})\approx0.380 - 0.641 p;\\
Re(\omega_{s=1,l=2})\approx-0.202 + 1.279 p,
\nonumber\\
Im(\omega_{s=1,l=2})\approx0.46133 - 0.73422 p;\\
Re(\omega_{s=1,l=3})\approx-0.406 + 1.965 p,
\nonumber\\
Im(\omega_{s=1,l=3})\approx0.474 - 0.753 p.
\end{eqnarray}
\end{subequations}

In formulas (\ref{scalarline}) and (\ref{elmline}), the linear approximation for $l=s=0$ for the scalar and  $l=s=1$ for the electromagnetic field, is carried out for values of $p$ which are larger than $1.04443$. It is possible that we do not observe the strict linear behavior at small $l$ and relatively small values of $p$, because the WKB methods is less accurate in this regime. This is also clearly visible in figures \ref{ris:one}a and \ref{ris:two}a.
\begin{table}[ht] \label{table2}
\begin{center}
\begin{tabular}{ | p{3cm} |  p{3cm} | p{3cm} |p{3cm} |p{3cm} |}
\hline
$p(\alpha$) & WKB 3-orders & WKB 4-orders &WKB 5-orders &WKB 6-orders\\
\hline
\multicolumn{5}{|c|}{Re$(\omega)$} \\
\hline
0.96943 & 1.10646 & 1.10827 & 1.10827 & 1.10840 \\
0.99443 & 1.14405 & 1.14621 & 1.14621 & 1.14637 \\
1.01943 & 1.18157 & 1.18415 & 1.18414 & 1.18432 \\
\hline
\multicolumn{5}{|c|}{Im$(\omega)$} \\
\hline
0.96943 & -0.25916 & -0.25874 & -0.25872 & -0.25869 \\
0.99443 & -0.27844 & -0.27791 & -0.27788 & -0.27784 \\
1.01943 & -0.29841 & -0.29776 & -0.29772 & -0.29768 \\
\hline
\end{tabular}
\caption{\label{tab:tabletwo}The dependence of the real and imaginary parts of the quasinormal frequencies for the scalar field ($s=0$) on the dimensionless parameter $p$ for $l=2$.}
\end{center}
\end{table}
\begin{table}[ht] \label{table3}
\begin{center}
\begin{tabular}{ | p{3cm} |  p{3cm} | p{3cm} |p{3cm} |p{3cm} |}
\hline
$p(\alpha)$ & WKB 3-orders & WKB 4-orders & WKB 5-orders & WKB 6-orders\\
\hline
\multicolumn{5}{|c|}{Re$(\omega)$} \\
\hline
0.96943 & 1.49941 & 1.49954 & 1.49954 & 1.49955 \\
0.99443 & 1.54815 & 1.54905 & 1.54905 & 1.54907 \\
1.01943 & 1.59725 & 1.59833 & 1.59833 & 1.59836 \\
\hline
\multicolumn{5}{|c|}{Im$(\omega)$} \\
\hline
0.96943 & -0.26254 & -0.25417 & -0.25419 & -0.25419 \\
0.99443 & -0.27267 & -0.27252 & -0.27255 & -0.27254 \\
1.01943 & -0.29165 & -0.29145 & -0.29149 & -0.29149 \\
\hline
\end{tabular}
\caption{\label{tab:tablethree}The dependence of the real and imaginary parts of the quasinormal frequencies on the dimensionless parameter $p$ for the electromagnetic field ($s=1$), $l=3$.}
\end{center}
\end{table}
The values obtained in table \ref{tab:tableone} and graphically presented in figure \ref{ris:one} have a significant difference from the data provided in the work \cite{Cai:2015fia}. The reason of this discrepancy between our results and the data presented in \cite{Cai:2015fia} is probably related to potentially insufficient accuracy when integrating the black hole solution: the time domain integration  \cite{Gundlach:1993tp} used in \cite{Cai:2015fia} requires  very high precision for discrete data for $A(r)$ and $B(r)$. In our case, it is examined an analytical metric, which makes it possible to improve the accuracy of the calculations. The accuracy of our calculations is confirmed by the smoothness of $\omega$ as a function of $p$ and reproducing the Schwarzschild limit when $p\rightarrow p_{min}$. Indeed, for example, for $p_{min}$ (given by [\ref{p_min}]), $l=2$ and $r_{0}=2$ we have $\omega =0.483642-0.096766i$ which is very close to the Schwarzschild limit $\omega =0.4836-0.0968i$ \cite{Konoplya:2003ii}. In our case, the function $\omega(p)$ tends to be linear, while in the case with the data obtained in \cite{Cai:2015fia}, the dependence of $\omega$ on $p$ is not smooth (see fig. \ref{ris:one}), which clearly indicates an unacceptably large numerical error in their calculations. In addition, the data presented in \cite{Cai:2015fia} does not allow one to see that the Schwaraschild limit can be reproduced as the mode with the smallest $p$ is rather far from the Schwarzschild mode. The correctness of our results is also indicated by ``the convergence'' of the obtained values for the frequency at higher WKB-orders. That is demonstrated in table \ref{tab:tabletwo} and in table \ref{tab:tablethree}.

\subsection{Frequency dependence of quasinormal modes with the mass-term} \label{sec:mass}
Let us consider the effective potential for a massive scalar field:
%
\begin{eqnarray}
V_s(r)=\dfrac{l(l+1)A(r)}{r^2}+\dfrac{B(r)\dfrac{\partial A}{\partial r}+A(r)\dfrac{\partial B}{\partial r}}{2r}+\mu^2 A(r).
\end{eqnarray}

We consider how the behavior of quasinormal modes changes with the increase of the mass-term. Let us fix $p = 1$ and $l=20$, so that the spectrum is in its eikonal regime and can be described by the WKB approximation very well. We find values for the quasinormal frequencies of this effective potential. Recently it has been shown that strictly speaking the WKB formula cannot describe the regime of quasiresonances accurately \cite{Konoplya:2017tvu} but can clearly show the tendency of the imaginary part to approach zero. Then, the extrapolation of the WKB data will indicate the existence of quasiresonances. We choose, as an example, $p = 1$, $l = 20$.\\
\begin{figure}[ht]
\vspace{-4ex} \centering \subfigure[]{\includegraphics[width=0.4\linewidth]{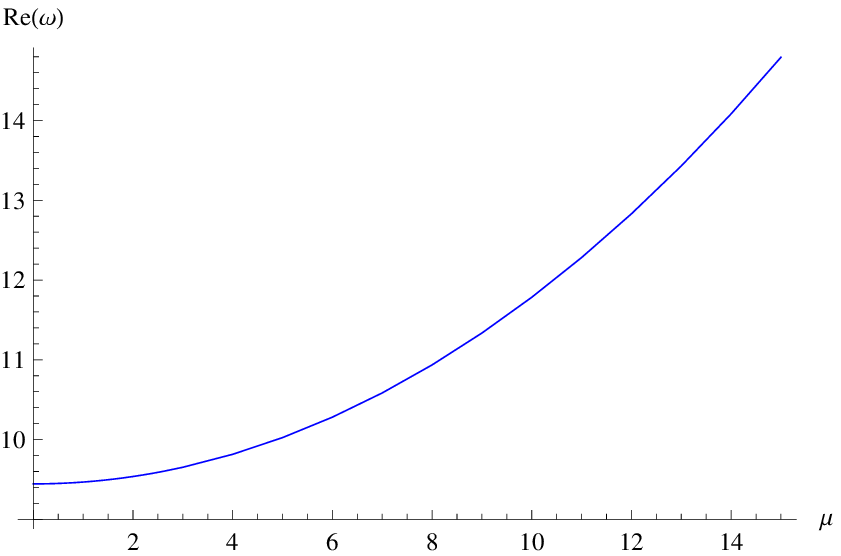} \label{fig:actuatorscouplingSheme_aaaa} }
\hspace{4ex}
\subfigure[]{
\includegraphics[width=0.4\linewidth]{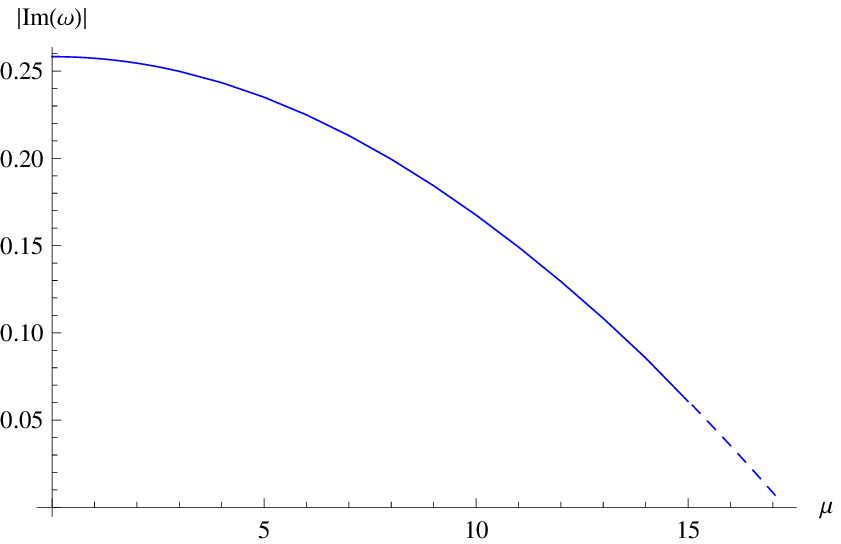} \label{fig:actuatorscouplingSheme_bbbb} }
\caption{ Dependence of frequency components on mass $\omega(\mu)$ with $s=0$, $l=20$, $p=1$, $\mu=0..16$:
 \subref{fig:actuatorscouplingSheme_aaaa} , Re$(\omega)$;
 \subref{fig:actuatorscouplingSheme_bbbb} , absolute value of Im$(\omega)$.}
 \label{ris:three}
\end{figure}

As can be seen in figure \ref{ris:three}, the real oscillation frequency increases with increasing field mass. At the same time, the damping rate of the oscillations tends to zero smoothly. This suggests that mass modes are scattered more slowly than massless modes and, under certain conditions, for massive fields, the oscillations become undamped, which are quasi-resonances.

\section{Conclusions}\label{sec:conclusions}

We considered quasinormal modes of the black hole in the quadratic gravity. The dependence of the modes on the new parameter $p$ was determined for different values ​​of the angular parameter $l$. From figures \ref{ris:one} and \ref{ris:two} it is obvious, that $l$ is increasing, then the real and imaginary parts of $\omega$ tend to the linear dependence on $p$. It is found that the quasinormal modes of the non-Schwarzschild black hole solution are characterized by larger real oscillation frequency and larger damping rate than the Schwarzschild  branch. In \ref{sec:mass} we added a massive term to the test scalar field. For such a field, for sufficiently large masses, the existence of undamped oscillations is shown, which are quasiresonances discussed in the earlier literature for Schwarzschild \cite{Konoplya:2004wg}, Reissner-Nordst\"{o}rm \cite{Ohashi:2004wr} and Kerr solutions \cite{Konoplya:2006br}. We have also shown that the previously published results of \cite{Cai:2015fia} for quasinormal modes of a massless scalar field are inaccurate.

It would be also interesting to consider perturbations of the gravitational field in the same black hole background. Using the WKB approach Hawking radiation for the above non-Schwarzschild black hole can  be considered via solution of the classical scattering problem.

\acknowledgments{
The author acknowledges Roman Konoplya for suggesting this problem and critical reading of the manuscript. This paper was supported by the Research Center for Theoretical Physics and Astrophysics, Faculty of Philosophy and Science of Sileasian University at Opava.}
\newpage

\begin{appendix}
\section{Quasinormal frequencies obtained with the help of the sixth WKB order formula}

\begin{table}[ht] \label{table1}
\centering
\begin{tabular}{ | p{1.4cm} | p{0.6cm}| p{2cm} | p{2cm} |  p{0.6cm}| p{2cm} | p{2cm} |}
\hline
 \multirow{2}{*}{$p(\alpha)$}
 & \multicolumn{3}{c}{$s=0$}
 \vline
 & \multicolumn{3}{c}{$s=1$}
 \vline\\
 \cline{2-7}
  & l & Re$(\omega)$ & Im$(\omega)$ & l & Re$(\omega)$ & Im$(\omega)$ \\
\hline
0.89443 & 0 & 0.21198 & -0.23715 & 1 & 0.50737 & -0.19556 \\
        & l & 0.60208 & -0.20741 & 2 & 0.93952 & -0.20142 \\
        & 2 & 0.99480 & -0.20548 & 3 & 1.35021 & -0.20285 \\
0.91943 & 0 & 0.21089 & -0.28137 & 1 & 0.52196 & -0.21007 \\
        & l & 0.62411 & -0.22460 & 2 & 0.97267 & -0.21755 \\
        & 2 & 1.03257 & -0.22249 & 3 & 1.40006 & -0.21932 \\
0.94443 & 0 & 0.21742 & -0.31928 & 1 & 0.53603 & -0.22506 \\
        & l & 0.64591 & -0.24285 & 2 & 1.00562 & -0.23432 \\
        & 2 & 1.07045 & -0.24023 & 3 & 1.44986 & -0.23644 \\
0.96943 & 0 & 0.22794 & -0.35228 & 1 & 0.54957 & -0.24048 \\
        & l & 0.66746 & -0.26215 & 2 & 1.03833 & -0.25166 \\
        & 2 & 1.10840 & -0.25869 & 3 & 1.49955 & -0.25419 \\
0.99443 & 0 & 0.24059 & -0.38201 & 1 & 0.56255 & -0.25627 \\
        & l & 0.68877 & -0.28249 & 2 & 1.07075 & -0.26957 \\
        & 2 & 1.14637 & -0.27784 & 3 & 1.54907 & -0.27254 \\
1.01943 & 0 & 0.25444 & -0.40966 & 1 & 0.57496 & -0.27242 \\
        & l & 0.70984 & -0.30388 & 2 & 1.10282 & -0.28802 \\
        & 2 & 1.18432 & -0.29768 & 3 & 1.59836 & -0.29148 \\
1.04443 & 0 & 0.26918 & -0.43563 & 1 & 0.58671 & -0.28892 \\
        & l & 0.73071 & -0.32634 & 2 & 1.13453 & -0.30701 \\
        & 2 & 1.22222 & -0.31821 & 3 & 1.64737 & -0.31102 \\
1.06943 & 0 & 0.28128 & -0.46691 & 1 & 0.59804 & -0.30518 \\
        & l & 0.75115 & -0.34940 & 2 & 1.16570 & -0.32655 \\
        & 2 & 1.25986 & -0.33945 & 3 & 1.69591 & -0.33118 \\
1.09443 & 0 & 0.29754 & -0.49013 & 1 & 0.60835 & -0.32238 \\
        & l & 0.77165 & -0.37406 & 2 & 1.19656 & -0.34660 \\
        & 2 & 1.29760 & -0.36138 & 3 & 1.74429 & -0.35187 \\
1.11943 & 0 & 0.31242 & -0.51613 & 1 & 0.61802 & -0.33954 \\
        & l & 0.79188 & -0.39955 & 2 & 1.22688 & -0.36720 \\
        & 2 & 1.33510 & -0.38405 & 3 & 1.79155 & -0.37102 \\
\hline
\end{tabular}
\caption{\label{tab:tableone}Values of the quasinormal frequencies obtained with the help of the sixth WKB order formula (\ref{wkb}) with $n=0$, $r_{0}=1$ ; $l=0,1,2$ for $s=0$ and $l=1,2,3$ for $s=1$.}
\end{table}
\end{appendix}

\end{document}